\documentclass[pre,amsmath,amssymb,floatfix,twocolumn,eqsecnum]{revtex4}
\usepackage{graphicx}
\usepackage{times}
\usepackage{xcolor}

\begin{document}
\title{The hidden role of coupled wave network topology on the dynamics of nonlinear lattices}

\author{Sophia R. Sklan}
\affiliation{Department of Mechanical Engineering, University of Colorado Boulder,
Colorado 80309 USA}
\author{Baowen Li\footnote{e-mail: Baowen.Li@Colorado.EDU}}
\affiliation{Department of Mechanical Engineering, University of Colorado Boulder,
Colorado 80309 USA}

\begin{abstract}
In most systems, its division into interacting constituent elements gives rise to a natural network structure.
Analyzing the dynamics of these elements and the topology of these natural graphs gave rise to the fields of (nonlinear) dynamics and network science, respectively.
However, just as an object in a potential well can be described as both a particle (real space representation) and a wave (reciprocal or Fourier space representation), the ``natural'' network structure of these interacting constituent elements is not unique.
In particular, in this work we develop a formalism for Fourier Transforming these networks to create a new class of interacting constituent elements $-$ the coupled wave network $-$ and discuss the  nontrivial experimental realizations of these structures.
This perspective unifies many previously distinct structures, most prominently the set of local nonlinear lattice models, and reveals new forms of order in nonlinear media.
Notably, by analyzing the topological characteristics of nonlinear scattering processes, we can control the system's dynamics and isolate the different dynamical regimes that arise from this reciprocal network structure, including the bounding scattering topologies.
\end{abstract}

\maketitle
\section{Introduction}
The study of transport within nonlinear lattices is a seminal problem in computational physics, thermal transport, and applied math \cite{FPU1,FPU2,FPU3,FPU4,FPU5,FPU6,KAM,Heat,Heat2,Heat3,Nlin,KAM2,KAM3}.
Despite the variety of researchers in the field and the role of this topic as the origin of computational physics, many out-standing problems remain \cite{FPU3,FPU4,FPU5,FPU6}.
This problem is in part compounded by the discrepancies between the most commonly considered models.
While a great many nonlinear lattices have been considered for various application (including the $\phi^4$, Toda, FPUT (Fermi-Pasta-Ulam-Tsingou), FK (Frenkel-Kontorova), and discrete sine-Gordon lattices \cite{FPU1,Phi4,Toda,Toda2,Toda3,FK,FK2,FK3,FK4,DSG}), they are each adapted to some special case.
The absence of a unifying framework of nonlinear lattices means that it is not always simple to find comparable lattices to help explain the origin of the phenomena observed in these structures \cite{FPU6,KAM,Soliton,Soliton2,Soliton3,Eqp,Eqp2,Eqp3,Chaos,Phononics}.
This problem is compounded when the results of the nonlinear lattice toy models are applied to understand experimental data, as without a proper understanding of the origins of the nonlinear transport effects it is difficult to say how precisely any given analogy between a toy model and an experimental system.
As such, the widening of the set of possible nonlinear lattices to beyond these special cases is likely to extend the utility of this field to new classes of problems.

In this paper, we show how the previously considered nonlinear lattices contain an implicit network structure which constrained the potential anharmonic couplings to a narrow set of options.
By modifying this implicit network we can construct a variety of novel nonlinear lattices and use them to elucidate the connection between the form of the network and wave dynamics in a nonlinear lattice.
In particular, we show how these new topologies can give rise to greater control of the dynamics of individual modes and provide bounds on the possible dynamics of the lattice.
Finally, we conclude by examining potential realizations of these modified networks.

\section{Coupled Wave Network Formalism}
To begin, let us consider the implicit network structure of an arbitrary linear system divided into $N$ interacting elements.
Assuming that it can be represented through simple Hamiltonian dynamics, we parameterize the system as
\begin{equation}
m_\alpha\ddot{u}_\alpha=-F_\alpha(u_\alpha)+\underset{\beta}{\sum}k_{\alpha\beta}A_{\alpha\beta}(u_\beta-u_\alpha)
\end{equation}
where $m_\alpha$ is the effective mass of the $\alpha^{th}$ element of the system, $u$ is an arbitrary variable defining the state of a given site (typically displacement from equilibrium for mechanical systems), $F$ is a so-called onsite potential that measures self-interaction force (typically zero for mechanical systems with translational symmetry), $k$ is the effective inter-site interaction force, and $A$ is the adjacency matrix.
The adjacency matrix will play a special role in our formalism, so it is worth emphasizing that whenever two elements are coupled (interacting), $A_{\alpha\beta}=1$ and whenever they are decoupled $A_{\alpha\beta}=0$.
Moreover, for every adjacency matrix we can define a network.
Each row or column corresponds to a vertex of a graph and each non-zero value of $A$ corresponds to an edge (if $A\ne A^T$ the edges are directed and define a digraph).
When $k$ is non-uniform, the product $k_{\alpha\beta}A_{\alpha\beta}$ defines a weighted graph.

Assuming that $F$ is also linear, this system can be diagonalized by taking a Fourier Transform to give the spectrum of eigenmodes $\tilde{u}(\vec{q})$ with eigenvalues $\omega_{\vec{q},p}$ (where $\vec{q}$ is the wave vector and $p$ is the branch number of the eigenmode $-$ to simplify some later notation we define $\omega(-\vec{q})\equiv-\omega(\vec{q})$ which is valid because $\omega$'s sign is arbitrary).
In particular, when these equations represent a mechanical system, the eigenmodes are mechanical waves (phonons) that propagate through the system without interacting, i.e. a phonon gas.
If either $F$ or $k$ became nonlinear, though, this would introduce anharmonic couplings between the phonons, and imply that these waves are no longe the eigenmodes of the system.
Without loss of generality, we consider a nonlinear coupling of the form:
\begin{gather}\label{eq:real}
m_\alpha\ddot{u}_\alpha=-F_\alpha(u_\alpha) \\ +\underset{\beta}{\sum}A_{\alpha\beta}\left[k_{\alpha\beta}\pm\gamma_{\alpha\beta}(u_\beta-u_\alpha)^{s-1}\right](u_\beta-u_\alpha). \nonumber
\end{gather}
When we Fourier Transform this equation, the result (suppressing branch index) is
\begin{gather}
\ddot{\tilde{u}}(\vec{q})=-|\omega(-\vec{q})\omega(\vec{q})|\tilde{u}(\vec{q})\\+\gamma^\prime(\vec{q})\omega(-\vec{q})\underset{f=1}{\overset{s-1}{\prod}}\left[\underset{\vec{q}_f}{\sum}\omega(\vec{q}_f)\tilde{u}(\vec{q}_f)\right]\delta\left(\vec{q}-\underset{f}{\sum}\vec{q}_f\right)\nonumber
\end{gather}
where $\gamma^\prime=i^{s}\mathcal{F}[\gamma/m]/(N)^{(s-1)/2}$ is the renormalized Fourier Transform of the nonlinear coupling and the delta function for a system with periodic boundary conditions is defined as
\begin{equation}\label{eq:mod}
q_i-\underset{f}{\sum} q_{f,i}=0 \:\mathrm{mod}\: N
\end{equation}
(i.e. within a reciprocal lattice vector for each component of $\vec{q}$).
Now this equation can be rewritten into a more evocative form:
\begin{equation}\label{eq:recip}
\ddot{\tilde{u}}(\vec{q})=-|\omega(-\vec{q})\omega(\vec{q})|\tilde{u}(\vec{q})+\underset{\vec{q}_1}{\sum}A(\vec{q},\vec{q}_1)W(\vec{q},\vec{q}_1)\tilde{u}(\vec{q}_1)
\end{equation}
where
\begin{gather}
W(\vec{q},\vec{q}_1)\equiv\gamma^\prime(\vec{q})\omega(-\vec{q})\omega(\vec{q}_1) \\ \times\underset{f=2}{\overset{s-1}{\prod}}\left[\underset{\vec{q}_f}{\sum}\omega(\vec{q}_f)\tilde{u}(\vec{q}_f)\right]\delta\left(\vec{q}-\vec{q}^\prime-\underset{f}{\sum}\vec{q}_f\right).\nonumber
\end{gather}
Comparing equations \ref{eq:real} and \ref{eq:recip} reveals a fundamental correspondence between a set of real space elements with self-interaction $F/m$ and nonlinear coupling $(k+\gamma(\Delta u))/m$ and a set of reciprocal space elements with self-interaction $|\omega(-q)\omega(q)|$ and nonlinear coupling $W(q,q^\prime)$.
Most significantly, in each case the coupling term is proportional to an adjacency matrix $A$, meaning that in both real and reciprocal space we can define network structures.
In the mechanical model, the real space graph defines a series of masses (vertices) coupled by nonlinear springs (edges) whereas the reciprocal space graph defines a series of linear waves (vertices) coupled by nonlinear scatterings (edges).
It is this latter representation of the system that we term a Coupled Wave Network (CWN) which is formally equivalent to taking a Fourier Transform of a dynamical system's graph.
Notably, any combination of nonlinear scatterings (i.e. any sum over Feynman diagrams) is equivalent to taking a directed walk within the CWN.
As we shall now confine our attention to reciprocal space, tildes on $\tilde{u}$ shall be suppressed to emphasize the similarity between the real and reciprocal space networks.

The representation of nonlinear wave scattering as a network was been independently developed in continuous media \cite{DR1,DR2,DR3,DR4,DR5}.
However, due to the continuity of the systems they studied, the correspondence between real and reciprocal space network representations was not noted in these works.
Similarly, note that this formalism is distinct from other uses of networks in physics, such as the representation of entanglement in tensor networks \cite{TensNetw} or the representation of real space interactions, such as in Ref. \cite{MasterNetw}.
In the former, networks are used as a graphical representation of entangled many-body states as a convenient computational tool.
As in Feynman diagrams, every vertex in the network represents a tensor coupling multiple wave functions (edges).
In the latter, the network serves as a generalization of a free-body diagram, showing how disparate points in real space interact with each other.
In contrast to these two approaches, every vertex in our CWN is a wave and every edge is an interaction.

\begin{figure}
\begin{center}
\includegraphics[scale=0.42]{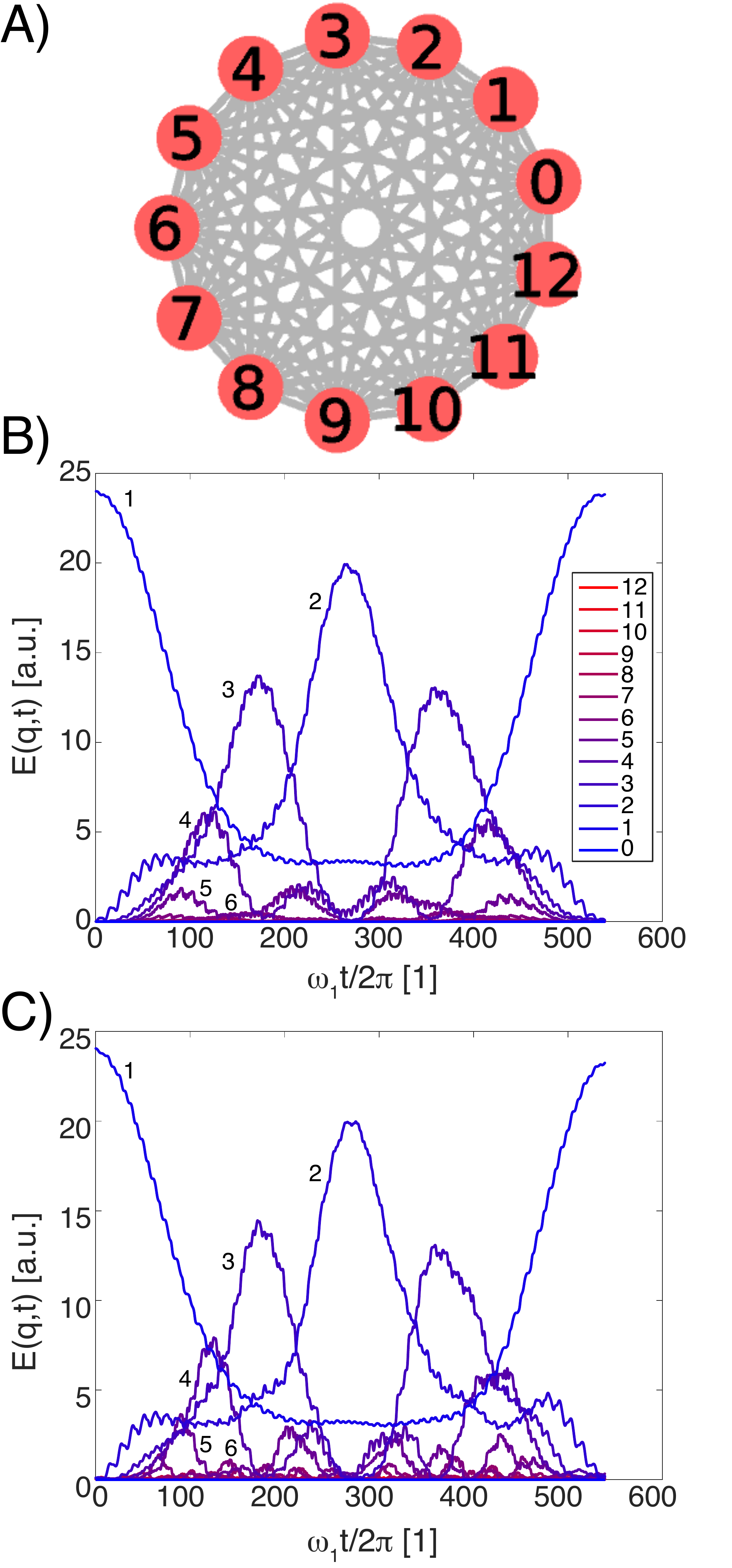}
\caption{\label{fig:FPU_0}(a) CWN structure of the $N=24$ FPUT-$\alpha$ lattice. Each vertex labeled by its corresponding $|q|$, where reciprocity allows the folding of $q<0$ vertices into $q>0$ vertices. (b) Harmonic energy of each mode. Since the system is reciprocal $u(-q,t)=u(q,t)$. Color indicates mode number. A coherent phonon population is initially injected in the $|q|=1$ modes and the system is left in isolation. As time progresses, anharmonic interactions cause the energy to flow and excite the $|q|>1$ modes. This excitation shows a clear hierarchy, with higher energy modes only excited once the lower energy ones reach a critical amplitude. As in the standard FPUT paradox, this excitation of the higher energy modes is clearly oscillatory, with energy flowing out of the $|q|>1$ modes and back into the $|q|=1$ modes after a critical period (the recurrence time). (c) Same as (b) with $0< k_BT\ll E_{coh}$
}
\end{center}
\end{figure}

While this derivation reveals the existence of CWNs, we must still determine its form.
From equation \ref{eq:mod} we see that valid (crystal momentum conserving) forms of phonon scattering are governed by modular arithmetic.
Since modular addition defines a group and every group is closed by definition, there must be a combination of ancillary modes that links any two given modes for an arbitrary strength FPUT-$\alpha$ coupling.
Thus an edge must exist between any pair of modes and the FPUT-$\alpha$ lattice is equivalent to a complete graph (CG) CWN (a network where every vertex shares an edge with every other vertex, see Fig. \ref{fig:FPU_0}a).

To validate our claim that the anharmonicity employed in equation \ref{eq:real} does not affect the generality of our derivation, let us consider the following variations.
A similar derivation holds for anharmonic on-site potentials (except for the elimination of factors of the frequency in the anharmonicity term).
Moreover, as this relation holds true for every integer rank anharmonicity and the addition/strengthening of links to a CG is still a CG, any anharmonicity with a valid Taylor expansion is still a CG.
Nor does the spatial dimension affect this, since each basis vector in the Brillouin zone obeys the same modular addition constraint.
Furthermore, adding a harmonic spatial modulation to the anharmonicity only shifts the modular addition constraint from equaling 0 to equaling the modulation's wavelength, which still defines a group and still gives a CG CWN.
Modifying the harmonic lattice structure would shift the values of $\omega_q$ but leaves the CWN constraint unchanged.
Changing boundary conditions would affect the CWN constraint, but as the most common conditions are either periodic or isolated (which is still a modular arithmetic constraint, albeit a more complex one \cite{ModalFPU1,PR}), the CG CWN appears to be a common feature of the most popular nonlinear lattice models.
Thus we see that the most common nonlinear lattice models all correspond to the same CWN, with their only difference being the strength of the couplings in this topology.

Given the universality of the CG CWN, we shall focus our attention on the FPUT lattice and further specialize to the one-dimensional FPUT-$\alpha$ lattice, where $s\equiv3$.
The motif (the CWN element corresponding to a single coupling term) associated with a nonlinearity of rank $s$ is an $s$ vertex complete sub-graph, so the FPUT-$\alpha$ lattice is associated with triangle motifs.
While the dynamics of such a lattice are quite well studied \cite{FPU3}, it will be an important reference and as such we reproduce the standard reciprocal space dynamics in Fig. \ref{fig:FPU_0}b.

Note that the dynamics of this simulation appear significantly noisier than the standard simulation, as the reciprocal space representation of the equation of motion has different error terms than the real space representation \cite{ModalFPU1} (indeed, real space simulations with lower accuracy show no such noise).
Rigorously, we can be sure that this noise is a numerical artifact as it produces fluctuations in the total energy at a characteristic frequency greater than the highest frequency phonon mode.
By confining our attention to qualitative aspects of the system which operate at a longer time scale than these fluctuations, we can ensure that this noise does not affect the validity of our qualitative results.
For quantitative results, we make use of the separation of time scales between the exact result and the numerical artifact noise and filter out the Fourier components corresponding to these fluctuations via a numerical low-pass filter (the modification of our simulations to facilitate this filtering are discussed more fully in Sec. \ref{sec:rand}).
This filtering removes almost all of the total energy fluctuation, ensuring that our results are both quantitatively and qualitatively rigorous.

We generate these (and all subsequent) numerical results using a 24 site FPUT-$\alpha$ lattice with $m=k=a=1$, $\gamma=1/4$, initial conditions of
\begin{gather}
u(q,0)=\frac{-i}{\omega(q)}\sqrt{E_{coh,q_0}\delta(||q|-|q_0|)+2k_BTn_B(q,k_BT)},\nonumber\\
\dot{u}(q,0)=0 
\end{gather}
(where $E_{coh}$ is the energy of an initially coherent population at $|q_0|$, $k_BT$ is the temperature, $\omega_q=\omega_0\sin(\pi q/N)$, $\omega_0=2\sqrt{k/m}$, and $n_B$ is the classical Boltzmann distribution $|\omega(q)|\mathrm{exp}[-\omega^2(q)/k_BT]/\Sigma_q|\omega(q)|\mathrm{exp}[-\omega^2(q)/k_BT]$), and numerical integration is performed via the Dormand-Price adaptive step-size Runge-Kutta 4-5 algorithm (which maintains fixed accuracy by modifying the size of the time steps).
Integration was carried out for one recurrence time of the FPUT-$\alpha$ lattice.
To match the initial conditions of the real space standard FPUT-$\alpha$ dynamics $E_{coh}=2N$, $q_0=1$, and $k_BT=0$ unless otherwise noted. 

One important feature of the FPUT-$\alpha$ lattice is that energy flows through a specific series of modes when spreading out from the initial coherent excitation \cite{FPU2,PR}.
These initially excited modes on this relaxation pathway are particularly important, as they serve as the gateways that limit the flow of energy to the entire population \cite{CNT1,CNT2}.
When the thermal energy is negligible compared to the coherent excitation, this gateway mode is twice the frequency of the coherent mode, as second harmonic generation (SHG) dominates (for a generic FPUT lattice of rank $s$, the dominant process is $s-1$ HG).
For an exact picture of these dynamics, see the video of the FPUT-$\alpha$ CWN's dynamics in the online supplement.

\section{Tailored CWN Topology}
While the previously studied nonlinear lattices have been constrained to be CGs in the CWN framework, other topologies are possible by relaxing the constraint of short-range (local) interactions.
In this section we show how the modification of the CWN away from a CG creates new opportunities to control the dynamics of specific modes.

\subsection{Gateway Control}\label{sec:GC}

\begin{figure}
\begin{center}
\includegraphics[scale=0.42]{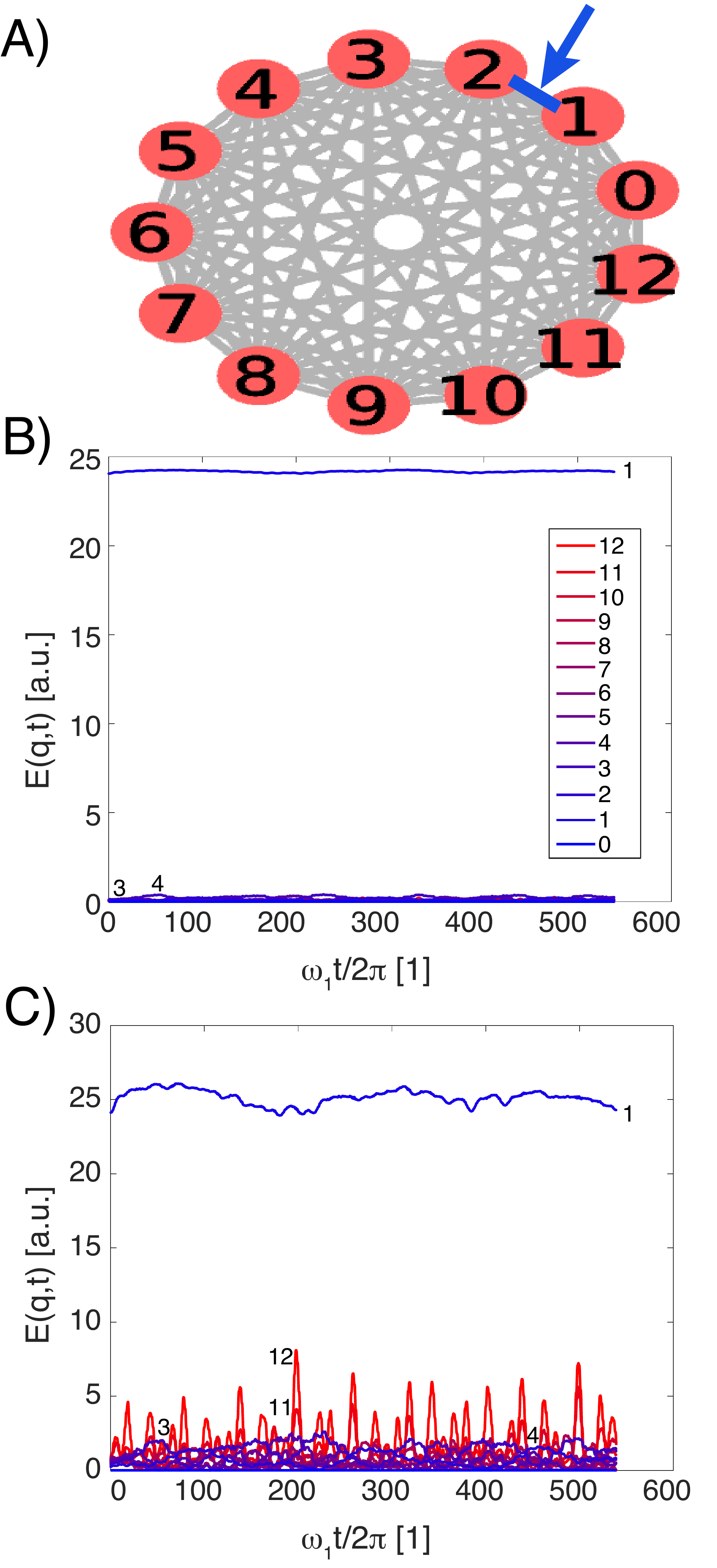}
\caption{\label{fig:gateway}FPUT-$\alpha$ lattice modified to remove the first gateway coupling mechanism. (a) CWN structures, removed edges are highlighted in blue and marked by an arrow. (b) Dynamics at $k_BT=0$ (c) Dynamics at $0<k_BT\ll E_{coh}$.
}
\end{center}
\end{figure}

We begin by returning to the observation that ended our previous section $-$ that SHG is the dominant gateway mode for an initially coherent phonon population.
To illustrate this effect we consider an FPUT-$\alpha$ lattice with a coherent excitation in the $|q|=1$ modes and remove the SHG coupling that links them to the $|q|=2$ modes (Fig. \ref{fig:gateway}a).
Simulations at $k_BT=0$ (Fig. \ref{fig:gateway}b) show no excitation of the other modes and the system appears entirely harmonic.
This is to expected though, as while the SHG coupling scales like $E_{coh}$, the other coupling modes scale like $\sqrt{E_{coh}k_BT}$, and should therefore be trivial at $k_BT=0$.
To detect the effect of modifying the CWN, then, we increase the temperature to $k_BT=E_{coh}/32$ and repeat the simulations for a modified FPUT-$\alpha$ (Fig. \ref{fig:gateway}c) and regular FPUT-$\alpha$ (Fig. \ref{fig:FPU_0}c).
Comparing these two figures reveals a reduction of the excitation of higher order modes by at least a factor of two (and up to an order of magnitude) due to the suppression of SHG for the fundamental mode.
(The $|q|=2$ mode in particular never exceeds its initial thermal energy, as it only acts as a donor to the higher order modes.)
As total energy is conserved, this also implies that significantly more energy is stored within the fundamental mode than the higher order modes, especially when compared to the typical FPUT dynamics (energy loss decreased by over an order of magnitude).
So by merely cutting a single edge on the CWN (suppressing a single scattering pathway), we have dramatically reduced the decay rate of the coherent excitation (even at non-zero temperature).

\subsection{Pathway Engineering}\label{sec:PE}
\begin{figure}
\begin{center}
\includegraphics[scale=0.3925]{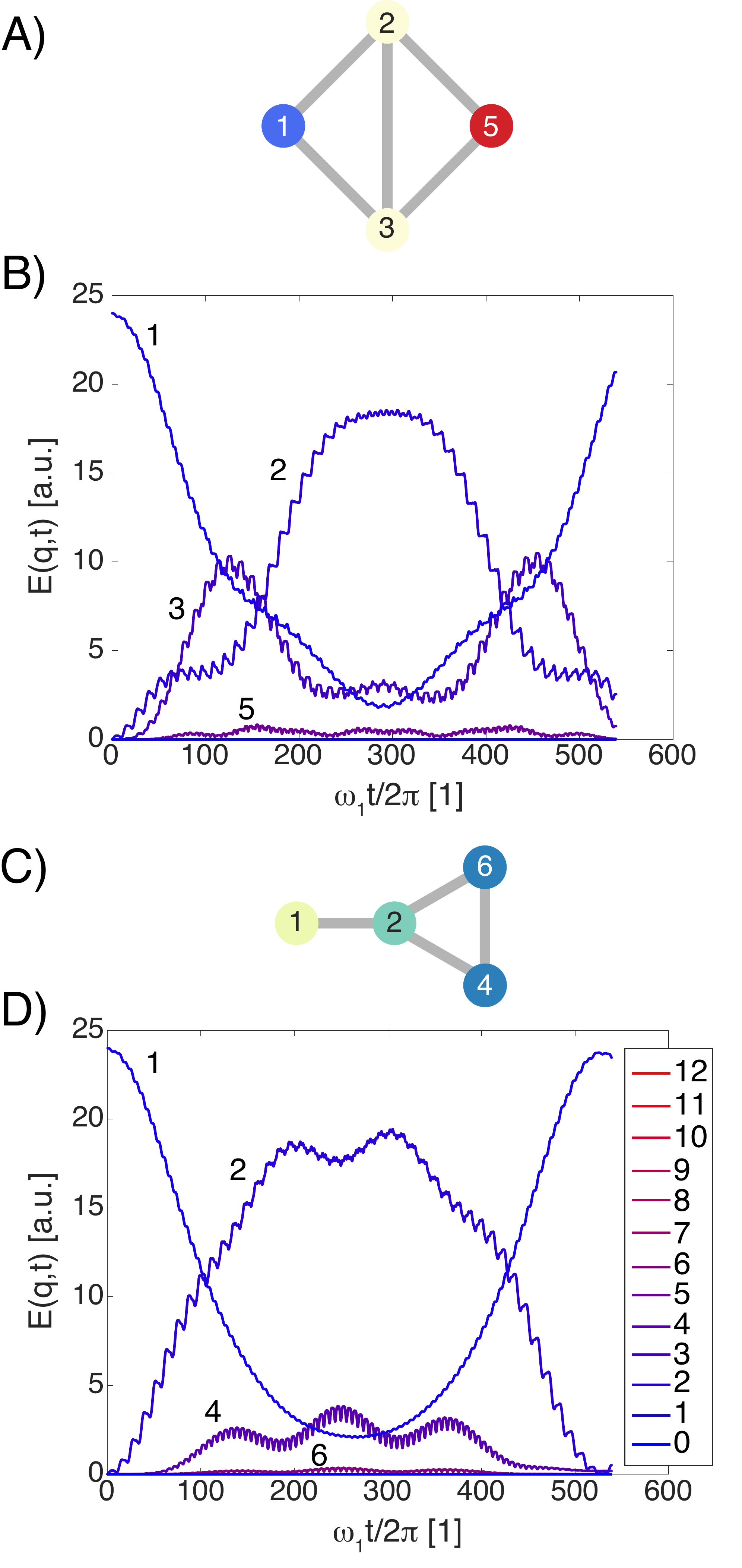}
\caption{\label{fig:eng}Engineered relaxation pathways in the modified FPUT-$\alpha$ lattice. (a) CWN of the relaxation pathway engineered to excited the $|q|=2,3,5$ modes, other modes trivially coupled and omitted. (b) Corresponding dynamics. (c) CWN of the relaxation pathway engineered to excited the $|q|=2,4,6$ modes, other modes trivially coupled and omitted (d) Corresponding dynamics.
Amplitude of the $|q|=6$ mode is small but non-zero, unlike the inactive modes where $E(q,t)$ is strictly 0 at all times.
}
\end{center}
\end{figure}

The ability to control the dynamics of specific modes by eliminating certain scattering pathways can extend beyond the simple elimination of the first gateway mode of the previous subsection.
Instead, by cutting specific subgraphs of the CG CWN, we can tailor the combination of activated modes.
That is, by only allowing subgraphs that incorporate specific parts of the standard relaxation pathway, we can direct the ordered excitation of an arbitrary combination of modes.
So, for example, the excitation of the $|q|=2$ modes via SHG from $|q|=1$ can lead to the excitation of the $|q|=3$ modes.
Via the modular arithmetic that govern mode conversion, similar rules exist for other combinations of modes or coherent excitations of modes other than the $|q|=1$ ones.
To illustrate this possibility, we engineer two different relaxation pathways and use them to tailor the excitation of specific phonon modes.
In the first (Fig. \ref{fig:eng}a), we can excite $|q|=2,3,5$ from the initial excitation (Fig. \ref{fig:eng}b), while in the second (Fig. \ref{fig:eng}c) we excite $|q|=2,4,6$ (Fig. \ref{fig:eng}d).
Note that, beyond the necessity of sharing the initial coherent population and SHG pathway ($|q|=1,2$) modes, these two pathways lead to the excitation of completely different phonon populations.
Note also that this is distinct from the existence of the even mode sub-manifold of the FPUT lattice \cite{PR}, as the initial excitation of $|q|=1$ should lead to the excitation of all other modes in the standard FPUT lattice (Fig. \ref{fig:FPU_0}b).
The sub-manifolds considered in Ref. \cite{PR} are much less stable than the engineered excitation pathways that we consider here, as the sensitivity to initial conditions that those sub-manifolds possess is eliminated through the tailoring of the allowed phonon couplings.

\subsection{Effects of Topological Distance}
\begin{figure}
\begin{center}
\includegraphics[scale=0.42]{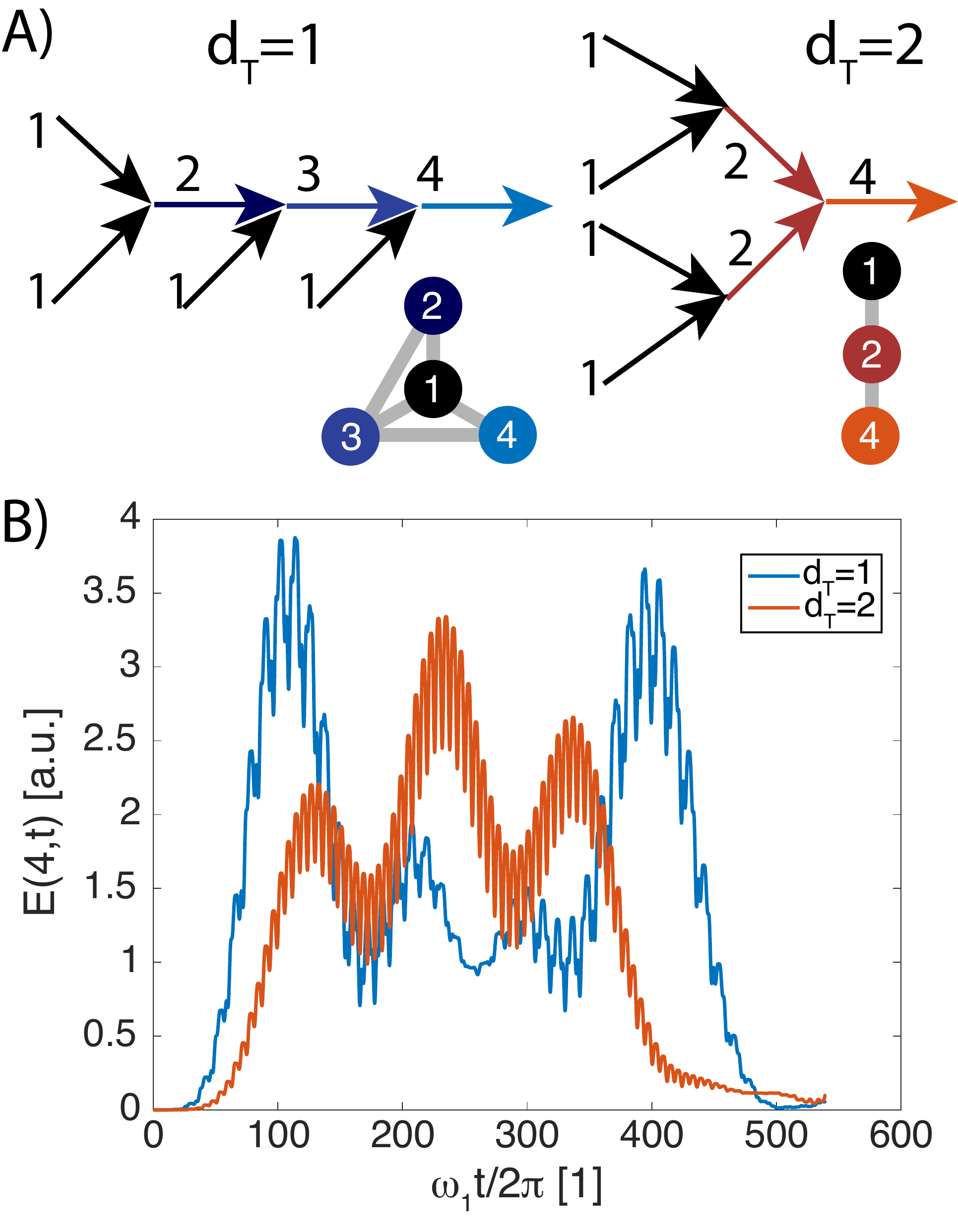}
\caption{\label{fig:TopoDist} Effects of topological distance on mode dynamics. (a) Feynman diagram (top) and CWN (bottom) representation of the $|q|=1\to|q|=4$ excitation pathway for topological distance 1 (left) and 2 (right). (b) Energy dynamics of the $|q|=4$ mode for the $d_T=1$ (blue curve) and $d_T=2$ (orange curve) topologies.
}
\end{center}
\end{figure}

As a final consideration of how the form of the CWN affects the dynamics of the excited phonons, we shall consider the effect of vertex-vertex distance in the CWN (or topological distance, to differentiate it from real space separation).
By topological distance we refer to the minimum separation between vertices in our network, in particular their separation from the initial coherent excitation at $|q|=1$.
Since the further a mode is from the initial excitation the more scattering will be required the excite it, this topological distance might appear to be equivalent to the number of scattering events.
However, this is not the case, which we can show by contrasting the dynamics of the $|q|=4$ mode when excited by two different scattering pathways (Fig. \ref{fig:TopoDist}a).
While the degree of nonlinearity is the same in both pathways, the first always involves the inclusion of a new $|q|=1$ phonon and therefore is of topological distance one while the second includes only the products of the initial SHG scattering in its final scattering event and is therefore of topological distance two.
Contrasting the dynamics of these anharmonically generated $|q|=4$ phonons in Fig. \ref{fig:TopoDist}b reveals that their time dependence is qualitatively different too. 
The mean energy and recurrence time are greater in the topological distance one case and the two cases have distinct beat patterns.
Thus, while these two pathways are equivalent in terms of strength of anharmonicity, they are distinct in terms of dynamics and we can associate this distinction with their different topological distances.
Similar situations should obtain for other scattering pairs, although as we shall see in section \ref{sec:dist} the topological distance has very strong effects for higher topological distances that obscure the subtler effect that we observe here.

\section{Topological Characterization of Energy Transfer in Random CWNs}\label{sec:rand}
One advantage of the CWN framework over that of Feynman diagrams or scattering is that networks possess a number of new characteristics  to describe their properties.
This allows us to find new correlations between the response of nonlinear lattices and the topological properties of the CWN.
To do this, it is necessary to consider a great many networks with the same topological properties, rather than isolating particular topologies that reveal novel effects (as we did in the previous sections).
We shall consider two specific topological characteristics of wave coupling, the topological distance discussed previously and the weighted vertex degree (or vertex strength).
When calculating the mean of the harmonic modal energies of this section we shall filter out the high frequency fluctuations that violate energy conservation, as they are a numerical artifact.
Since this filtering can introduce ringing at discontinuities, for each random network we select an integration domain of that network's recurrence time (whereas previously we used the FPUT-$\alpha$'s recurrence time).

\subsection{Dependence on Topological Distance} \label{sec:dist}
To determine the correlations between topological distance and transfer of energy from the coherent excitation to other modes, we consider the time-average energy of each mode in thirty six random lattices.
These random lattices are constructed as reciprocal subgraphs of the CG CWN (i.e. the FPUT-$\alpha$ lattice), essentially constraining the subgraphs to complete sets of triangle motifs linking any six vertices related by reciprocity.
Reciprocal motifs were added to the graph randomly until all vertices were connected to ensure that the distance was well defined (by powers of the adjacency matrix).
To ensure that non-trivial dynamics would exist at $0K$, the SHG scattering of the fundamental mode was always included in the CWN (see \ref{sec:GC}).
Specific additional couplings would be included as well (since most relaxation pathways occupy only a small part of parameter space), together forming a seed CWN to which the random couplings would be added.

\begin{figure}
\begin{center}
\includegraphics[scale=0.215]{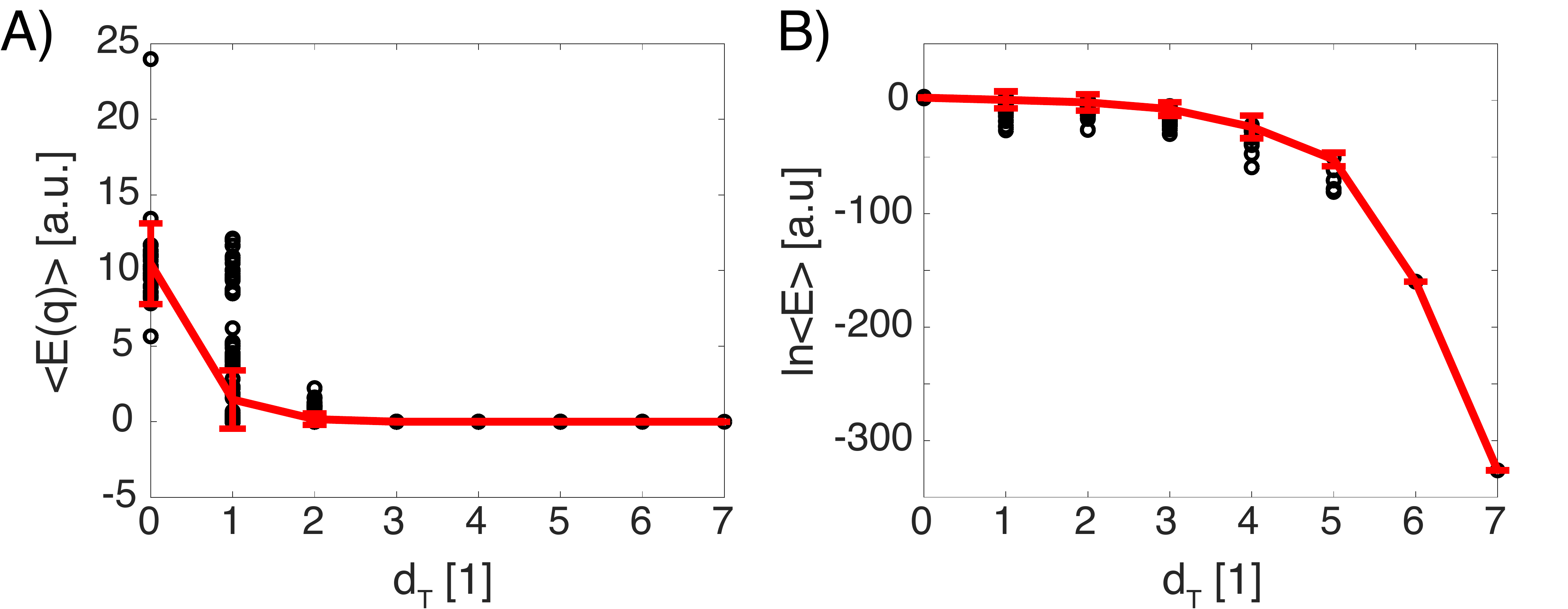}
\caption{\label{fig:dT}Mean harmonic modal energy in 36 random CWNs versus topological distance $d_T$ for (a) linear and (b) semilogarithmic plots. Points denotes individual modes while the red line indicates the mean value as a function of topological distance. Error bars for $d_T>2$ are smaller than the width of the curve for the linear plot and are absent for $d_T>6$ for both plots as only one data point exists for those distances.
}
\end{center}
\end{figure}

The results of these seeded random graphs are shown in Fig. \ref{fig:dT}a, where each point is the mean energy of a mode in a given graph.
Notice that mean amplitude decays quickly with distance, with only $d_T\le2$ showing any appreciable amplitude.
Using a semilogarithmic plot of this data (Fig. \ref{fig:dT}b) reveals that the energy decays approximately like $E(q,d)=E(q,0)\exp(-\xi(q)d^4)$.
Note also that there is a significant fraction of the $d_T=1$ domain that falls above the error bars. This is due to the FPUT-$\alpha$ relaxation pathway, as the presence or absence of couplings in the random lattices that fall along this pathway creates a bimodal distribution of amplitudes for $d_T=1$.
The comparatively smaller energy that above the error bars of $d_T=2$ also corresponds to this pathway, but since it requires active combinations of the $d_T=1,2$ pathways, it is comparatively rarer and weaker.
As for the coherent excitation itself ($d_T=0$), note that there is considerable variance around the mean value, almost as much as the $d_T=1$ relaxation pathway induced variance.
This is principally due to the conservation of energy within the lattice, as decreased (increased) conduction to the higher modes results in greater (lesser) energy confined to the fundamental mode.
This suggests a certain degree of tunability to the coherent population's thermal relaxation as a function of specific topologies, but it also reveals a very clear shielding of phonon energy transfer within the CWN structure.
Unlike in real space, where the FPUT lattice displays a simple power law scaling with the separation between sites \cite{AC1,AC2,AC3}, the energy in reciprocal space remains highly localized.
The FPUT-$\alpha$ lattice coupling, in fact, produces quite small energy flux for $d_T>2$, an effect that has been masked in other studies of the nonlinear lattice as the CG CWN necessarily has $d_T\le1$ for any pair of modes.
Given this clear restriction in energy transfer across topological distances and the approximately linear increase in mean topological distance with system size, we find that the effects that we have observed in the $N=24$ FPUT-$\alpha$ do not appreciably change when extended to larger systems.
This is particularly the case as random lattices are unlikely to contain more than a few elements of the relaxation pathway for large systems, so energy doesn't flow through more than the first handful of modes in the pathway.

\begin{figure}
\begin{center}
\includegraphics[scale=0.42]{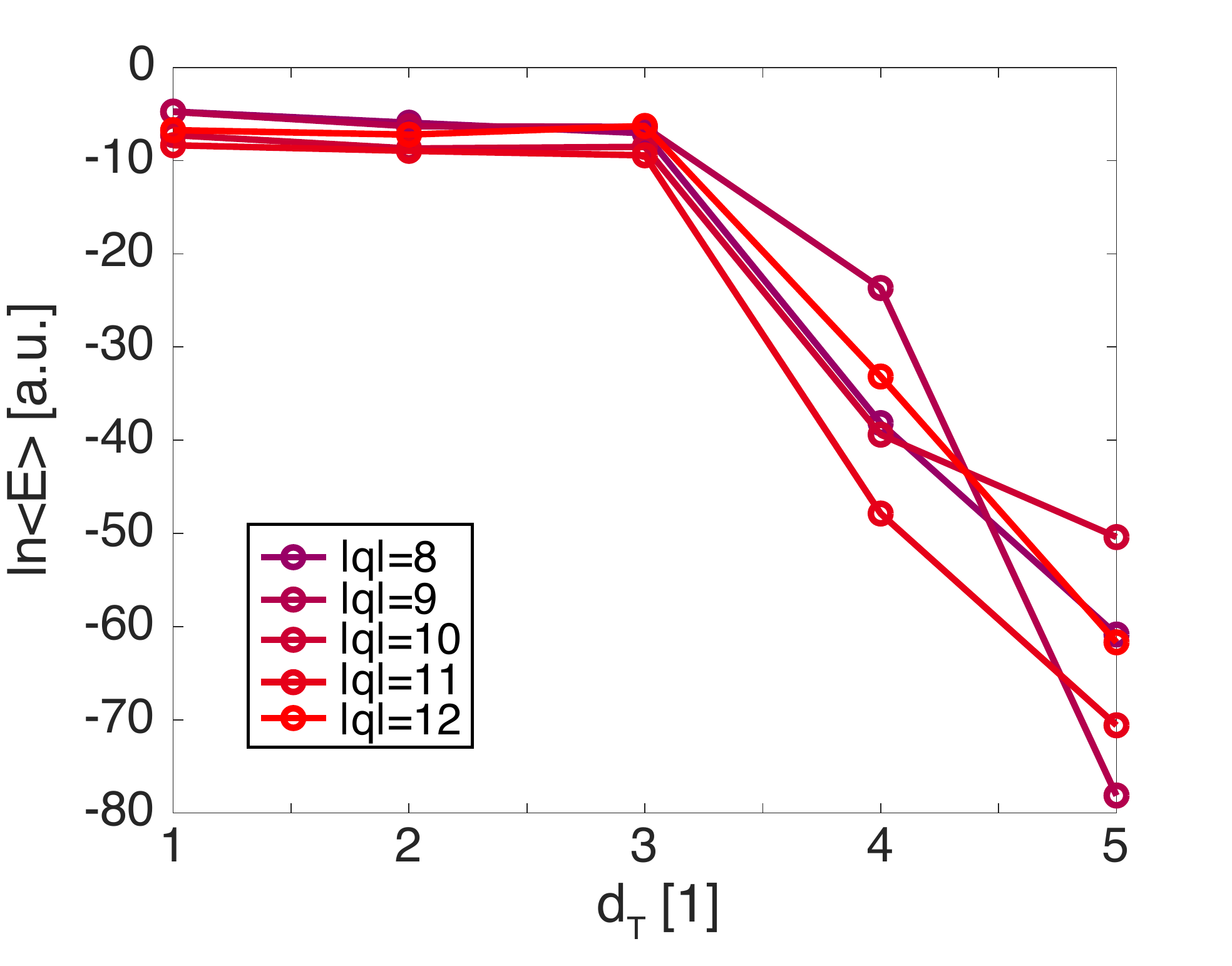}
\caption{\label{fig:NUtr}Transition from normal dominated to Umklapp dominated scattering as a function of topological distance from the coherent source in random CWN. Data points correspond to mean energy values of different random CWN with fixed $q$ and $d_T$. Note the clear cross-over at $d_T=3$, the critical distance for these modes.
}
\end{center}
\end{figure}

Examining the interaction between topological distance and mean energy of each mode reveals an interesting result $-$ the transition between normal and Umklapp scattering is reflected in this distribution.
Specifically, while both normal and Umklapp scattering can be present for modes connected below a critical distance ($d_{Tc}=\left\lfloor\ln q/\ln 2\right\rfloor$ for the FPUT-$\alpha$ lattice with the fundamental mode excited, see Sec. \ref{sec:app} for details), Umklapp scattering dominates above this critical distance from the coherent source.
This is clearly seen in Fig. \ref{fig:NUtr} that focuses on the set of modes in the $N=24$ lattice where $d_{Tc}=3$ (data for $d_{Tc}=2$ is similar, but the transition is obscured by the presence of only two data points in the normal regime).
This also partially explains the quartic scaling observed previously, which averaged over all mode numbers.
Since the normal dominated regime decays linearly and the Umklapp dominated regime decays quadratically, the quartic scaling is necessary to fit this discontinuity.

\subsection{Dependence on Weighted Degree}\label{sec:degree}
\begin{figure}
\begin{center}
\includegraphics[scale=0.2]{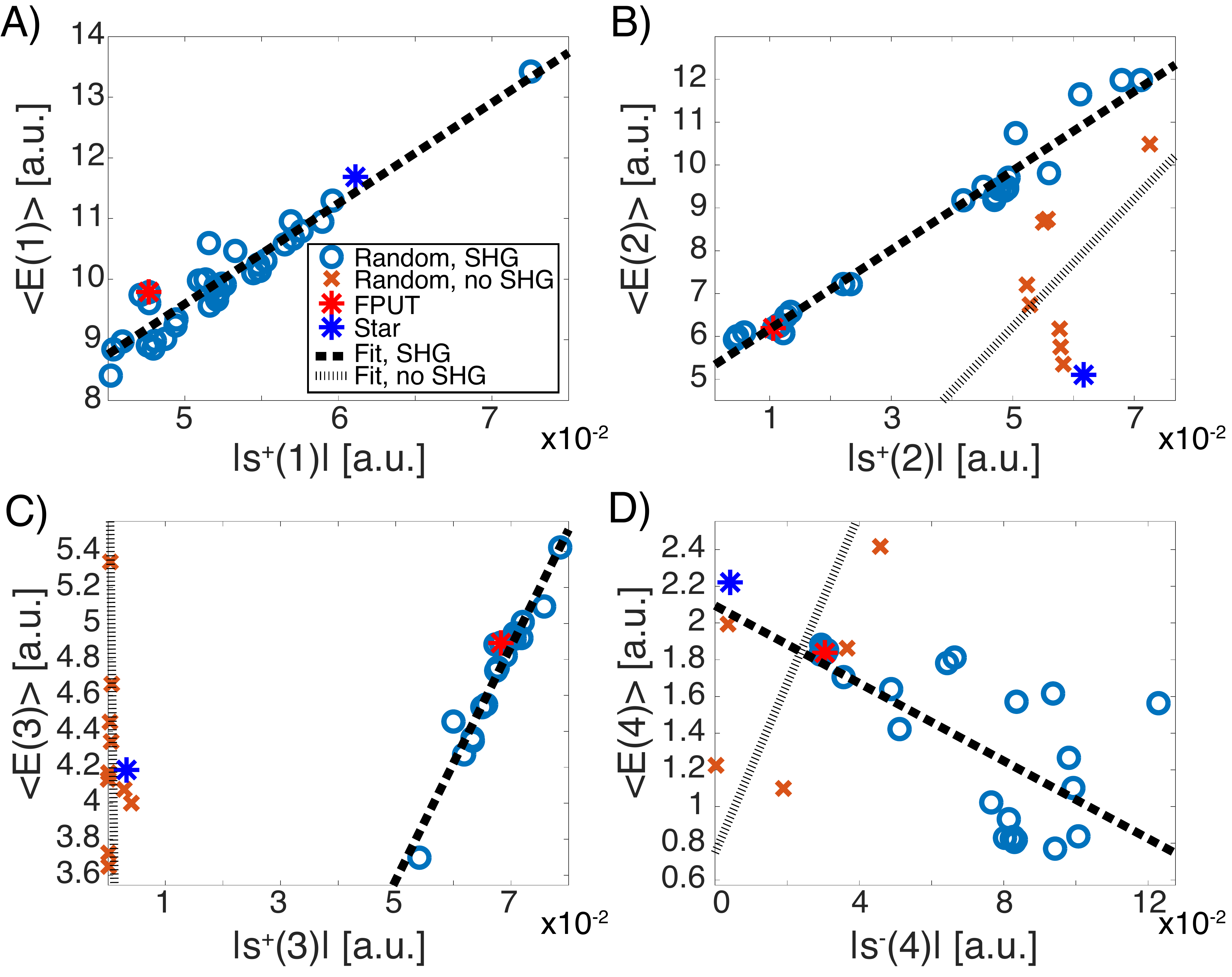}
\caption{\label{fig:degree}Mean energy versus absolute value of degree for the first four modes of random CWNs. Blue circles are CWNs with critical the SHG pathway present. Orange circles are CWNs with the critical SHG pathway absent. Black lines are the linear fits in each case. The red star is the FPUT-$\alpha$ graph and the blue star is the $|q|=1$ star CWN (Sec. \ref{sec:bound}). Neither star is included in the linear fitting. (a,b,c) $|s^+(q)|$ for $|q|=1,2,3$ respectively. (d) $|s^-(4)|$.
}
\end{center}
\end{figure}

Although we can use the topological distance to distinguish between normal and Umklapp dominated scattering, even at fixed $q$ the mean energy is not a simple monotonic function of topological distance.
This is in part due to topological distance being an unweighted measure of interaction strength, whereas paths in the CWN are not all equivalent.
Some, such as the relaxation pathways, will carry more energy than others.
As such, we turn to the weighted degree of each vertex as a more effective means of characterizing the interaction.
The weighted degree is given by the time average differential force between a vertex and all its neighbors.
For a reciprocal CWN, this takes the simplified form:
\begin{equation}
s_{i}=\underset{j}{\sum}|\omega_{q_i}\omega_{q_j}\omega_{q_{i-j}}|\langle \left[u(q_i)-u(q_j)\right]u(q_{i-j})\rangle
\end{equation}
where the brackets denote a time average.
Plotting the mean energy as a function of degree reveals the reason for this non-monotonic dependence between mean energy and topological characteristics (see Fig. \ref{fig:degree} for representative examples, others shown in the Sec. \ref{sec:app}) $-$ the presence or absence of specific paths in the CWN will shift these parameters.
Specifically, the presence or absence of an SHG pathway for each mode will determine how mean energy scales with topological distance (in the FPUT-$\alpha$ lattice).
For odd modes, it is the out-going SHG edge that is critical (as there is no in-coming edge for these modes), whereas for  even modes it is the in-coming edge that is critical (except for $|q|=2$, where the in-coming edge is present in all realizations of the random CWN, making the out-going edge critical).
Using this rule, we identify three regimes:
when the critical SHG coupling of a mode is present, the mean energy (not the log of mean energy) scales approximately linearly with weighted degree and both mean energy and mean degree are relatively high.
When the critical SHG coupling of a mode is absent, the mean energy scales approximately linearly with weighted degree and mean energy is relatively high while mean degree is relatively low.
When the coupling is Umklapp dominated, this distinction is relatively unimportant as mean energy and mean degree are both low and so no scaling is observed.

\subsection{Bounding Topologies of the CWN}\label{sec:bound}

\begin{figure}
\begin{center}
\includegraphics[scale=0.42]{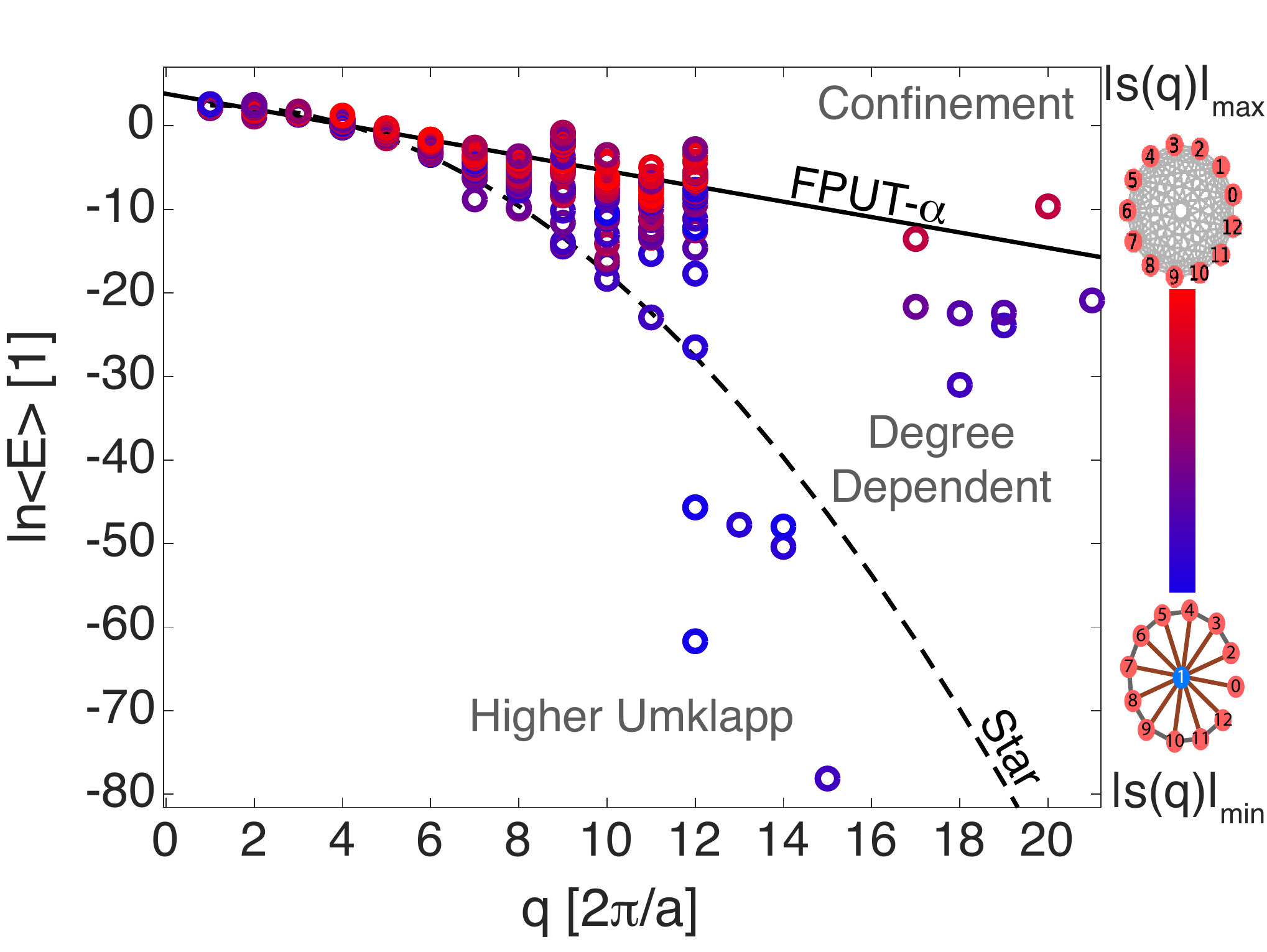}
\caption{\label{fig:2BZ}Log of mean energy versus wave vector for random seeded CWN topologies. Points correspond to individual realizations of the CWN, with color denoting the normalized absolute value of their weighted degree. Solid line denotes the linear FPUT-$\alpha$ bound on the CWN scaling, dashed line denotes the quadratic $|q|=1$ star graph lower bound on the CWN scaling. CWNs corresponding to these bounds are shown next to their color scale. Umklapp dominated modes have been corrected to the second Brillioun zone. 
}
\end{center}
\end{figure}

To conclude our analysis of the role of CWN topologies on seeded random couplings, we consider the role of mode number on this dependence.
In doing so we shall make the following corrections to our data.
First, any modes where Umklapp scattering dominates (i.e. above the critical distance, see Sec. \ref{sec:dist}), we shall shift the wave vector to the second Brillouin zone, as the phonon is effectively acting like this higher order mode.
We also exclude any data points where the mode is inactive (zero mean energy), since we are concerned with the $|q|$ versus $\ln\langle E\rangle$ dependence.
Combining these corrections gives us Fig. \ref{fig:2BZ}, where we see that the mean energy typically falls between two bounds.
On the upper bound is the FPUT-$\alpha$ graph, i.e. the maximum degree CWN.
Conversely, the lower bound corresponds to the $|q|=1$ star graph, where the only valid triads are those that contain the $|q|=1$ vertices, which forms the minimum non-trivial (unweighted) degree for the higher harmonics.
Between these two bounds we see a clear scaling with increasing weighted degree leading to increased mean energy, although this scaling is non-monotonic (see Sec. \ref{sec:degree}).
Note though that these bounds are not strict, we often observe data points above or below them.
For the points below these bounds, these correspond to topologies where the modes are Umklapp dominated.
While we cannot reconstruct the exact mode number corresponding to the scattering of a random graph, it is likely that this Umklapp scattering takes these modes out of the second Brillouin zone and into higher order zones.
Given that the FPUT-$\alpha$ graph scales linearly with $q$ and the star graph scales quadratically, there will necessarily be some correction where any data point would fall within their bounds.
For points above these bounds, conversely, this effect is real and rigorous.
Interestingly, the points far above the FPUT-$\alpha$ bounds tend to have lower weighted degree than those just below it.
In fact, these data points correspond to topologies where many modes are inactive and most of the energy is confined to a small number of active modes.
Since the total energy is conserved, this confinement necessarily results in modes with greater energy than the FPUT-$\alpha$ bound.

\section{Potential Experimental Realizations}
While the non-trivial CWNs introduced in this work reveal a great deal of novel effects in nonlinear lattice dynamics, the complexity of the interactions that they require will likely impede immediate experimental confirmation in nonlinear lattices.
In particular, each triad motif ($q_1-q_2-q_3=0\;\mathrm{mod}\;N$) corresponds to a real space coupling of the form
\begin{gather}
E\propto\underset{n}{\sum}\left((u_n-u_{n-1})\underset{\Delta_1}{\sum}(u_{n+\Delta_1}-u_{n+\Delta_1-1})e^{iq_1\Delta_1a}\right.\nonumber\\\left.\times\underset{\Delta_2}{\sum}(u_{n+\Delta_2}-u_{n+\Delta_2-1})e^{iq_2\Delta_2a}\right),
\end{gather}
i.e. a highly delocalized complex coupling between lattice sites.
While this can be simplified using special cases, e.g. the SHG only coupling is
\begin{gather}
E\propto\underset{n}{\sum}\Bigg((u_{n}-u_{n-1})\nonumber\\\left.\times\underset{\Delta}{\sum}(u_{n+\Delta}-u_{n+\Delta-1})(u_{n-\Delta}-u_{n-\Delta-1})\right)
\end{gather}
(which at least has the advantage of being purely real and symmetric), it is still a highly delocalized force.
This suggests that realizations of this system would require a rather precise ability to tailor long-range nonlocal interactions between lattice sites.
Such a requirement would likely be easiest to accomplish in atomic systems, particularly atoms in a lattice of optical traps, where there is a greater degree of control over inter-atomic couplings \cite{BEC1,BEC2,BEC3}.
While these systems work best for very small numbers of atoms, the limited topological distance of interactions in random CWNs (Sec. \ref{sec:dist}) suggests that this is an acceptable constraint.
For larger systems, however, there would need to be some form of additional coupling to create the long range delocalized interactions.
This could be accomplished by using internal stresses, as in the Kroner-Eringen nonlocal elasticity model \cite{KE1,KE2,KE3}, which would allow for couplings of the form
\begin{gather}
\sigma_{nonlocal}(q)=\lambda(q)\sigma_{local}(q)\\
\sigma_{nonlocal}(x)=\int\lambda(x-y)\sigma_{local}(y)dy
\end{gather}
for nonlocality parameter $\lambda$.
This suggests that the CWN could arise in other systems possessing effective fields, as in composite structures (fibers, laminates, etc.) \cite{KE4,KE5,KE6,KE7}, as the integration over internal structure can introduce these effective nonlocal interactions.
A particularly important class of materials where nonlocal mechanical effects are known is piezoelectrics \cite{PZ1,PZ2,PZ3,PZ4}, as the charge-induced long range interactions are easily tailored and the material itself is not particularly exotic.
Still, as nonlocal effects are not presently known for phonons (aside from the Kroner-Eringen model), it is likely that additional research into effective phonon-phonon couplings will be necessary before these numerical predications are testable.

On the other hand, many of these effects are not confined to solid mechanicals.
CWNs are already known to exist for nonlinear resonances in continuous media \cite{DR1,DR2,DR3,DR4,DR5}, making fluid dynamics a promising field for realizing these results.
The principal challenge there would be tuning the wave interactions in the fluid to realize specific CWNs.
Furthermore, effective nonlocal interactions can exist in complex networks, which encompass a broader set of systems than simple mechanical systems.
In particular, non-trivial CWNs are likely to exist within social networks, as the nonlocal forces that produce such graphs would correspond to interpersonal interactions mediated by (complex) social pressure dynamics.
The possibility of two agents having their interactions influenced by the state of third agent is fairly exotic in mechanics (i.e. three-body forces are rare) but relatively mundane a state of affairs in social interactions.

\section{Conclusions}
In this work we have shown how the anharmonic interactions that define scattering in nonlinear media implicitly create a reciprocal network structure that defines the wave mixing (i.e. a coupled wave network).
By modifying this coupled wave network structure to create nonlocal nonlinear lattices, we have shown how it becomes possible to precisely control the dynamics of the phonons within the nonlinear lattice.
In particular, we have demonstrated how simple modifications of the CWN structure could be used to change the relaxation rate and reduce energy flux between coherent phonon populations and a thermal background.
Moreover, directing the energy down specifically tailored relaxation pathways or controlling the topological distance between modes can be used to control the specific combination of modes excited by the coherent population and their resulting time dynamics.
And most significantly, we have found graph topology signatures of different transport regimes within nonlinear lattices, including CWNs that effectively bound arbitrary anharmonic scattering mechanisms.
The variety of new effects seen within CWNs suggests that they can become a powerful tool for understanding the origin of nonlinear transport within real systems with long-range interactions,  provide an invaluable comparison when seeking to understand the nature of transport within the more conventional nonlinear structures, and open alternative avenues for understanding dynamics on complex networks more generally.

\makeatletter 
\renewcommand{\thefigure}{S\@arabic\c@figure}
\makeatother
\setcounter{figure}{0}

\section{Appendix} \label{sec:app}
%
%
%

\subsection{Normal-Umklapp Transition}
\begin{figure}
\begin{center}
\includegraphics[scale=0.2]{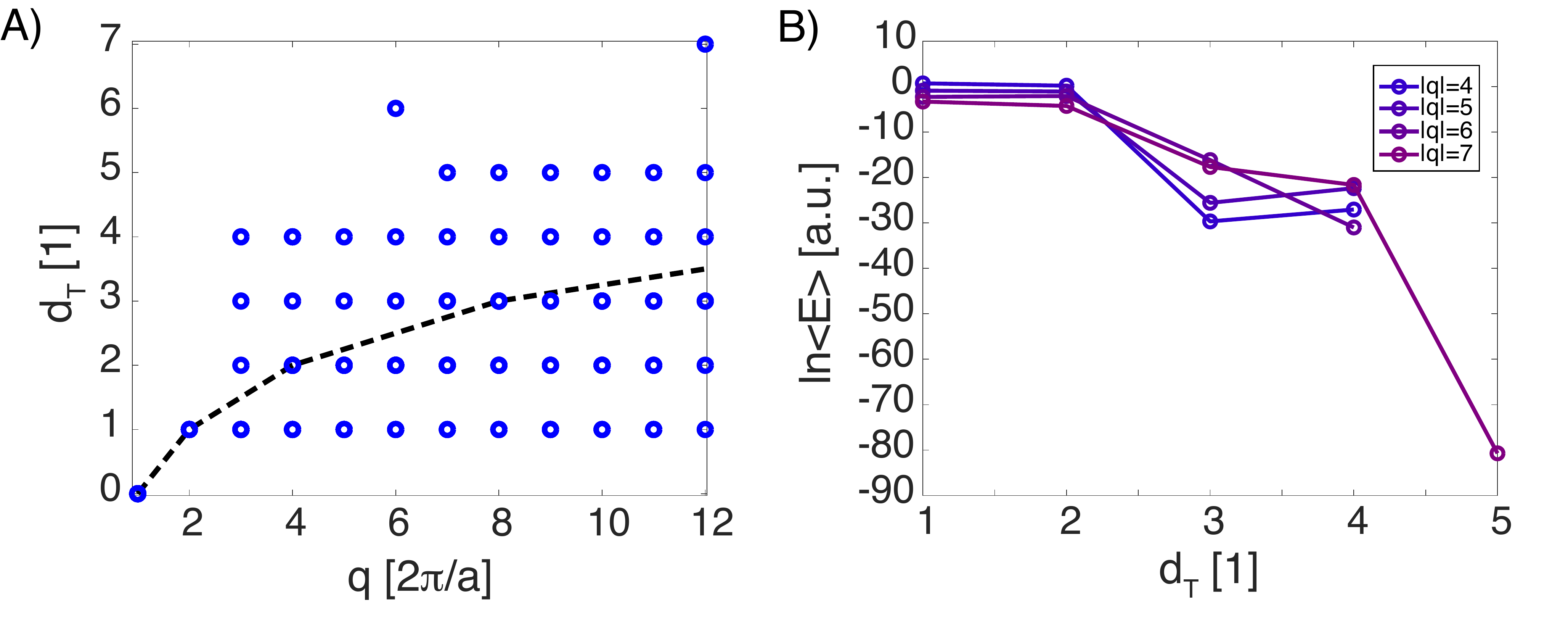}
\caption{\label{fig:NUtr2} (a) Distribution of topological distances versus wave vector realized in the seeded random CWNs. Circles denote the presence of a combination (ignoring multiplicity) and the dashed line denotes the critical threshold topological distance for a given wave vector. Points on or below the curve are normal scattering dominated, points above the curve are Umklapp scattering dominated. (b) Normal-Umklapp dominance transition for modes where $d_{Tc}=2$.
}
\end{center}
\end{figure}

Assuming no Umklapp scattering, the lowest order mode that can be generated at a given topological distance from a coherent source is equal to $s-1$ times the lowest order mode of the previous distance (for an anharmonicity of rank $s>2$).
Thus the maximum distance that a mode can be from a coherent source at $|q|=|q_0|$ is determined by the relation $|q|\le|q_0|(s-1)^{d_T}$, giving the critical threshold for $q_0=1$:
\begin{equation}
d_{Tc}=\left\lfloor\ln|q|/\ln(s-1)\right\rfloor.
\end{equation}
For $s=3$, this gives a scaling as shown in Fig. \ref{fig:NUtr2}a.
This behavior is reflected in Fig. \ref{fig:NUtr} and Fig. \ref{fig:NUtr2}b for $d_{Tc}=3,2$ respectively.

\subsection{SHG Dependence of Higher Harmonics}

\begin{figure*}
\begin{center}
\includegraphics[scale=0.2]{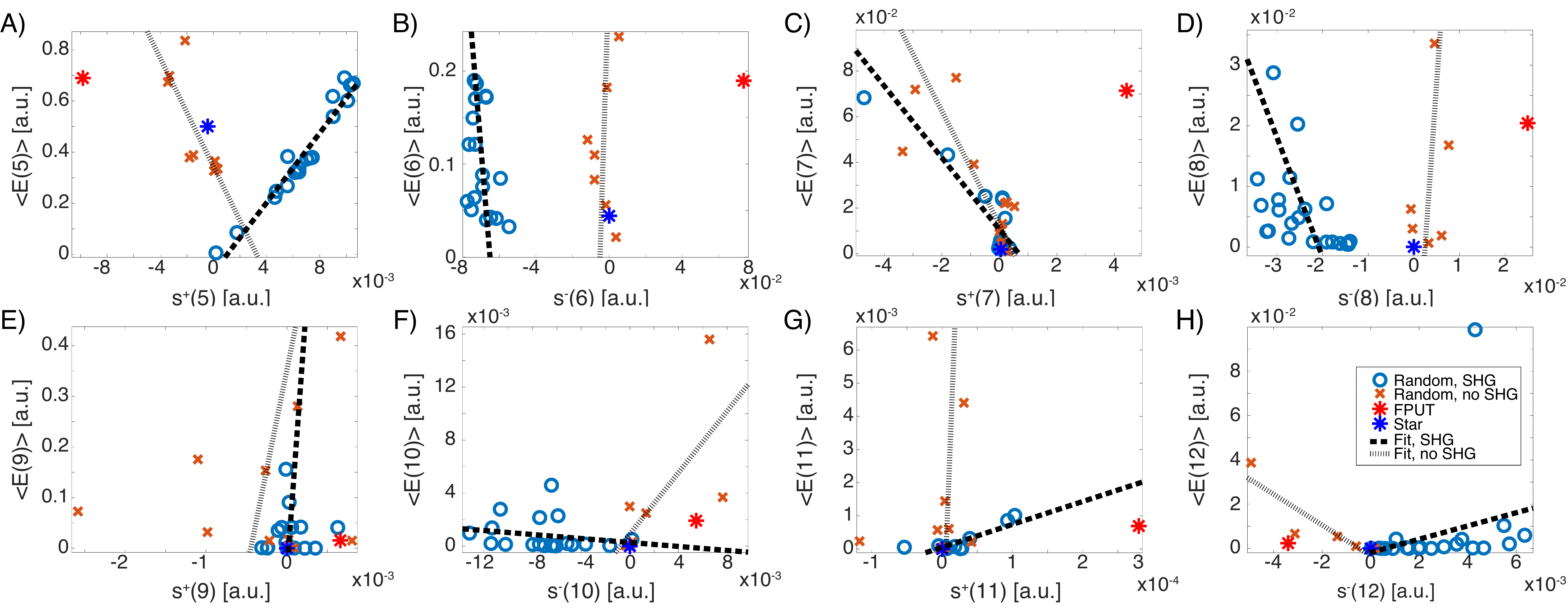}
\caption{\label{fig:deg2} Signed degree dependence of different modes ($|q|\in[5,12]$) for random seeded CWNs, following the same conventions as Fig. \ref{fig:degree}.
}
\end{center}
\end{figure*}

In Fig. \ref{fig:deg2} we continue plotting the degree dependence of different modes for random CWNs.
While the basic pattern is maintained from Fig. \ref{fig:degree}, certain features are worth emphasizing in the higher harmonics.
First, note that we have switched from plotting the absolute value of $s_i$ to its signed value.
While this somewhat obscures comparisons between the magnitude of $s_i$ and the different pathways explored in Sec \ref{sec:degree}, it reveals that the sign (i.e. the direction of mean force) is often different in these two cases.
Surprisingly, the sign of the FPUT-$\alpha$ lattice is often the same as the sign of the SHG absent path and opposite to the sign of the SHG present path, which in turn shares a sign with the star graph.
This is opposite to the trend of their magnitudes.
However, as the sign of $s_i$ only correlates with the imaginary component of the average differential nonlinear force, the role of $s_i$'s sign in CWN conduction remains somewhat obscure.

Additionally, the separation between the two pathways begins to break down at higher orders, particularly when the absolute value of $s_i$ is used.
This is in part due to the lower energy carried by these higher order modes, which causes the fluctuations from the random CWN topologies to have a stronger effect, but is also due to number of intermediate pathways that must be activated before the higher order modes can be generated.
This is likely why the odd modes typically display a weaker dependence on the presence of SHG pathways than the even modes (although the fact that the odd modes require an SHG out-going edge while the even modes require and SHG in-coming edge is another significant contributing factor).

\end{document}